\documentclass[11pt]{article}

\usepackage{amsmath,amssymb}
\usepackage{amsfonts}

\newcommand{\be}{\begin{equation}}
\newcommand{\ee}{\end{equation}}
\newcommand{\bea}{\begin{eqnarray}}
\newcommand{\eea}{\end{eqnarray}}
\newcommand{\eref}[1]{(\ref{#1})}

\newcommand{\BR}{\mathbb{R}}

\newcommand{\bra}[1]{\langle{#1}|}
\newcommand{\ket}[1]{|{#1}\rangle}
\newcommand{\vev}[1]{\langle{#1}\rangle}

\newcommand{\Ham}{{\widehat{H}}}
\newcommand{\al}{\alpha}
\newcommand{\om}{\omega}
\newcommand{\lam}{\lambda}
\newcommand{\ta}{\tilde{a}}
\newcommand{\Sig}{\Sigma_{t=0}}
\newcommand{\tpi}{\widetilde\pi}
\newcommand{\tphi}{\widetilde\phi}
\newcommand{\sqrtg}[1]{\sqrt{-g({#1})}}
\newcommand{\sqrtga}[1]{\sqrt{\gamma({#1})}}
\newcommand{\g}[1]{g_{#1}}
\newcommand{\tg}[1]{\widetilde{g}_{#1}}
\newcommand{\pig}[1]{\pi_g^{#1}}
\newcommand{\tpig}[1]{\tpi_g^{#1}}
\newcommand{\hatS}{\hat{S}}
\newcommand{\Sal}{\hat{S}_\alpha}
\newcommand{\meff}{m_{\rm eff}}
\newcommand\kk{\vec{k}}

\newcommand\x{\vec{x}}
\newcommand\y{\vec{y}}
\newcommand\nn{\nonumber}
\newcommand\ie{{\em i.e.}\ }
\newcommand\eg{{\em e.g.},\ }
\newcommand\cf{{\em cf.}\ }

\def\CD{{\cal D}}

\def\CH{{\cal H}}
\def\CI{{\cal I}}

\def\CO{{\cal O}}

\def\CU{{\cal U}}

\newcommand{\ds}[1]{{\rm dS}_{#1}}
\newcommand{\dS}[1]{{\rm dS}_{#1}}

\newcommand{\pa}{\partial}

\def\tr{\hbox{tr}}

\begin{document}

\rightline{hep-th/0406217}
\rightline{ITFA-2004-24}
\rightline{VPI-IPPAP-04-04}

\vskip 1.5 cm
\renewcommand{\thefootnote}{\fnsymbol{footnote}}
\centerline{\Large \bf Alpha-states in de Sitter space}

\vskip 1 cm
\centerline{{\bf 
Jan de Boer,${}^{1}$\footnote{\tt jdeboer@science.uva.nl}
Vishnu Jejjala,${}^{2}$\footnote{\tt vishnu@vt.edu} and
Djordje Minic${}^{2}$\footnote{\tt dminic@vt.edu}
}}

\vskip .5 cm
\centerline{${}^1$\it Instituut voor Theoretische Fysica}
\centerline{\it Valckenierstraat 65, 1018XE Amsterdam, The Netherlands}
\vskip .5 cm
\centerline{${}^2$\it Institute for Particle Physics and Astrophysics}
\centerline{\it Physics Department, Robeson Hall, Virginia Tech}
\centerline{\it Blacksburg, VA 24061, U.S.A.}

\setcounter{footnote}{0}
\renewcommand{\thefootnote}{\arabic{footnote}}

\vskip 1.5 cm
\begin{abstract}
Field theory in de Sitter space admits a one-parameter family of vacua determined by a superselection parameter $\al$.
Of these vacua, the Euclidean vacuum uniquely extrapolates to the vacuum of flat Minkowski space. 
States which resemble the $\al$-vacua can be constructed as excitations above the Euclidean vacuum.
Such states have modes $\al(k)$ which decay faster that $k^{(1-d)/2}$.
Fields in such states exhibit non-local correlations when examined from the perspective of fields in the Euclidean vacuum.
The dynamics of such entangled states are fully consistent.
If an $\al$-state with properties that interpolate between an $\al$-vacuum and the Euclidean vacuum were the initial condition
for inflation, a signature for this may be found in a momentum dependent correction to the inflationary power spectrum.
The functional formalism, which provides the tool for examining physics in an $\al$-state, extends to fields of other spin.
In particular, the extension to spin-2 may proffer a new class of infrared modifications to gravitational interactions. 
The implications of superselection sectors for the landscape of string vacua are briefly discussed.
\end{abstract}

\newpage

\section{Introduction}\label{sec:intro}

Cosmology provides a window into the Universe at early times, high energies, and large scales.
That in the beginning the Universe was very nearly homogeneous and isotropic is a fact adduced 
from observations of identical structures today in causally disconnected regions of spacetime.
The temperature of the cosmic microwave background radiation (CMBR), which maps out the surface
of last photon scattering, has an angular distribution that varies at the level of one part in 
ten thousand.
Small perturbations in the temperature are directly related to density perturbations at the time
of recombination, when the Universe first became transparent to radiation.
Large scale inhomogeneities in the Universe today (stars, galaxies, clusters, etc.) arise from
cosmological perturbations at early times.
Inflation provides a possible explanation for homogeneities and inhomogeneities in the density of matter
when photons decoupled \cite{infl}.
Recent astrophysical data \cite{wmap} provide corroboration for the inflationary paradigm. 

Understanding the large scale structure of the Universe today amounts to understanding in detail
the initial conditions that preceded the inflationary epoch.
During inflation, the large scale geometry is well approximated by a de Sitter spacetime, which
is the maximally symmetric solution to the vacuum Einstein equations with a positive cosmological 
constant \cite{desitter}.
A pure de Sitter phase for the early Universe is only an approximation, however.
The de Sitter geometry is empty and static.
It can model eternal inflation, but does not address such physically interesting phenomenological 
questions as the nature of the slow-roll cosmological phase transition and the turn-over to the
classical Friedmann-Robertson-Walker-Lema{\^\i}tre evolution.
Moreover, field theory in de Sitter space, even neglecting issues of back-reaction, is problematic
from the point of view of local observers.
The usual $S$-matrix intuition of local quantum field theory seems to fail, or rather is 
unsuitable \cite{ed}.

Quantum field theory in de Sitter space is further complicated by the existence of a one-parameter
infinite family of vacuum states consistent with CPT invariance \cite{motolla, allen}.
These vacua are labeled by a real number $\al$.
One of these vacua $\ket{\al = 0}$ is the unique vacuum that extrapolates to the standard Minkowski
vacuum in the limit where the cosmological constant vanishes (\ie the radius of curvature of the
de Sitter geometry becomes infinite) and has the same short distance singularities along the light 
cone as the usual flat space two-point function \cite{eucl, ghds, bd, burges, ratra}.
As well, the response function of an Unruh detector behaves thermally only in this vacuum \cite{burges},
which is called the Euclidean, adiabatic, or Bunch-Davies vacuum.
The other vacua $\ket\al$ are {\em formally} realized as squeezed states over the Euclidean vacuum 
\cite{qo}. 
This is only a formal correspondence because each of the vacua $\ket\al$ is the ground state of a 
different Hilbert space; $\al$ is a superselection parameter.
Understanding whether interacting field theory in any of the $\ket{\al\ne 0}$-vacua makes
sense as a consistent theory of physics is still a matter of debate \cite{glel}.
In this paper, we discuss a somewhat different scenario in which the notion of an $\al$-vacuum leads to 
meaningful new physics.

The beauty of inflation is that it is a future attractor.
The choice of initial conditions is somehow generic.
Many sets of initial conditions lead to the same observable astrophysics at late times.
The exponential expansion of the Universe during inflation washes out most of the details.\footnote{This
assumes that the language of effective field theory applies with the usual decoupling of scales.} 

Although there may be compelling arguments for preferring the Euclidean vacuum over the other
$\al$-vacua, there is no {\em a priori} justification for the belief that the Universe started out
as a fluctuation about the adiabatic ground state.
Rather, it is conceivable that the initial conditions were some fluctuation about some excited state.
If this excited state is in the same Hilbert space as the Euclidean vacuum (\ie it is a normalizable
excitation over the Euclidean vacuum), the physics inherits the field theoretic structure of the 
Euclidean ground state and there is no inconsistency with observation or need to employ non-standard
technology.
There are no unusual singularities in the Green function, for example, and the detector response
function is perfectly thermal.
Novel physics at late times can arise in such a scenario despite inflation if the excited state were
special.
We can consider physics in an $\al$-state, which is like the $\al$-vacuum in the infrared, but 
like the Euclidean vacuum in the ultraviolet. 
Such states are excited states in the same Hilbert space as the Euclidean vacuum and are perhaps
natural because, from the point of view of effective field theory, it makes no sense to demand
``alphaness'' to arbitrary precision; we expect new degrees of freedom (such as strings) to emerge 
at least near the Planck scale.
New physics is required to address the initial Big Bang singularity, or possibly, to fix pre-inflation
initial conditions.
Preparing an $\al$-state is fine-tuning, but this state is no more finely tuned than any other of
the manifold of conceivable initial conditions.

An $\al$-state is a squeezed state.
Given an $\al$-state as an initial condition, the field theory at late times will exhibit long-range
(Hubble scale) correlations.
Physics is non-local,
but the long-range correlations are simply a manifestation of entanglement in the initial state.
As no experiment has verified the locality of physics at Hubble distances, there is no inconsistency
with observation.

Modifications in the inflaton power spectrum, \ie the Fourier transform of the autocorrelation function
of the inflaton field, would be one potential signature for $\al$-like physics at early times.
If the initial condition was the Euclidean vacuum, the modes of a scalar field are thermally populated.
Inflation stretches the modes.
The thermally populated modes in the ultraviolet descend below the cutoff scale $\om$ that defines the 
effective field theory.
Certain initial conditions in the Euclidean vacuum lead to $\CO((H/\om)^2)$ deviations from the standard 
scale invariant spectrum \cite{stanford}.
Other authors consider initial conditions in an $\al$-vacuum and obtain $\CO(H/\om)$ deviations, which 
are potentially observable by the WMAP and Planck experiments as a signature of trans-Planckian physics,
although these results have engendered some controversy in the literature \cite{ulf, shiu}.
(See also Ref.\ \cite{lm}.)
The $\al$-states are one more set of initial conditions to probe primeval anisotropies in the CMBR. 

Another potential signature for new physics that is inspired by $\al$-states exists in the gravitational
sector.
Einstein's theory of gravity is a classical field theory valid at least in the
regime of distances larger than a few microns.
Modifications to the gravitational interaction at astrophysical scales are constrained by the systematics 
of galactic rotation curves, the Tully-Fisher luminosity relation, and the dynamics of clusters \cite{abfn}.
There is, however, no observational check to the General Theory of Relativity in the very deep infrared.
A number of scenarios have suggested various (possibly consistent) infrared modifications to the classical
theory of gravitation \cite{nima}.
If we treat fluctuations in the metric as being a spin-2 field in de Sitter space, then the technology of
$\al$-states for scalars will translate to a non-local redefinition of the spacetime metric.
Exploring this hypothesis is an interesting problem.

The organization of the paper is as follows.
In Section \ref{sec:dSft}, we shall briefly review scalar field theory in de Sitter space.
In Section \ref{sec:opal}, we discuss scalar fields in the operator formalism.
We shall explain how the $\al$-vacua arise from a Bogoliubov transformation of the Euclidean mode
functions and ladder operators and discuss the formal expression of the $\al$-vacua as squeezed
states over the adiabatic ground state.
We then present the construction of a normalizable squeezed state with $\al$-vacuum like properties
in the same Hilbert space as the Euclidean vacuum.
In Section \ref{sec:funal}, we revisit the definition of an $\al$-vacuum, this time in the functional
formalism. 
The wavefunctional for the Euclidean vacuum and the $\al$-vacua are constructed using the 
functional Schr\"odinger equation, which respects the de Sitter isometries, and the antipodal map,
$\hatS: \phi(x) \mapsto \phi(x_A)$, which sends a field at $x$ to the field at $x_A$, the point 
antipodal to $x$, in the de Sitter spacetime. 
The definition of the $\al$-vacuum wavefunctional is then modified to construct the wavefunctional
for a scalar field that is an excitation of an $\al$-state.
It is the physics of $\al$-states that we discuss in the functional formalism.

In Section \ref{sec:aldyn}, we consider two cases.
First, we suppose that $\al(k)$, the momentum space expression for the $\al$-state, is some fixed
profile.
In particular, we consider the dynamical evolution of a field $\tphi(x)$ in this state and compute its 
two-point function in the Euclidean vacuum.
If $\tphi(x)$ is the inflaton, the two-point function will imply a deviation from the predictions of standard 
inflationary cosmology.
Our results are consistent with momentum dependent corrections to the power spectrum that scale as
$\CO(H^2 \al(k))$, where $\al(k)$ specifies the initial condition.
Secondly, we make a first attempt at the more ambitious goal of treating $\al$ as a fully dynamical
parameter that interpolates between $\al$-vacuum-like behavior at early times and Euclidean 
behavior at late times.
In a heuristic sense, the ``vacuum'' will roll during inflation. 
We argue that the intuition we inherit from entangled states in Minkowski space makes such dynamics reasonable in
de Sitter space.
To make our discussion more precise, we construct a toy model of harmonic oscillator squeezed states and consider
the implications this has for the dynamics of $\al$-states in de Sitter space.

In Section \ref{sec:algr}, we comment on how the formalism we have developed applies to other 
dynamical degrees of freedom besides scalars.
In particular, we explore whether we can endow fluctuations in the background de Sitter metric with
$\al$-like character.
Section \ref{sec:conc} offers some concluding remarks.
We emphasize that superselection parameters must be part of any definition of the geography of the 
landscape of string vacua and comment upon some more general implications of superselection parameters in 
quantum gravity.

\section{Scalar Field Theory in de Sitter Space} 
\label{sec:dSft}

The maximally symmetric solution to the $d$-dimensional vacuum Einstein equations with constant,
positive curvature is de Sitter space.
The de Sitter geometry is realized as the hyperboloid
\be
\eta_{ab} X^a X^b = H^{-2} \equiv \ell^2 
\ee
embedded in $(d+1)$-dimensional Minkowski space, a construction which makes manifest the $SO(1,d)$
isometry group of $\dS{d}$.
The Hubble scale $H$ sets the cosmological constant $\Lambda = \frac{1}{2}(d-1)(d-2) H^2$
and the curvature $R = d(d-1)H^2$.
Spatial sections of the global manifold are spheres, $S^{d-1}$, but inflationary cosmology involves
planar sections.
A review of de Sitter geometry in various coordinate systems is contained in Refs.\ \cite{davies, review}.

Let us consider the free scalar field $\phi(x)$ in de Sitter space.\footnote{Our discussion extends to
free fields of other spins, as it depends only on properties of the de Sitter isometry group $SO(1,d)$.}
The symmetric Green function 
\be
G_\al(x,y) = \bra\al \left( \phi(x) \phi(y) + \phi(y) \phi(x) \right) \ket\al
\ee
in a de Sitter invariant state $\ket\al$ can depend only on the geodesic distance $d(x,y)$ \cite{ss, allen, allenj}:
\bea
d(x,y) = H^{-1} \cos^{-1} Z(x,y), && 
Z(x,y) \equiv H^2 \eta_{ab} X^{a}(x) X^{b}(y).
\eea
$Z(x,y)$ is defined in terms of the embedding space $\BR^{1,d}$.
Notice that $Z(x,x) = 1$ so that $d(x,x) = 0$ and that $Z(x,x_A) = -1$, where $x_A$ is the point antipodal
to $x$; $X^a(x_A) = - X^a(x)$ in terms of the embedding geometry.
(In global de Sitter space with spherical sections, the antipodal point to the north pole at time $t$ is
the south pole at time $-t$.)
Given the maximal symmetry of the de Sitter geometry, $G_\al(x,y)$ must be a function only of $Z(x,y)$:
$G_\al(x,y) = F(Z)$.
The Klein-Gordon equation for $G_\al(x,y)$ implies that \cite{allen, allenj} 
\be
\left( (Z^2-1) \frac{d^2}{dZ^2} + d Z \frac{d}{dZ} + m^2 H^{-2} \right) F(Z) = 0.
\label{hyper}
\ee
The solution to this differential equation is a hypergeometric function.
Given the invariance of the equation of motion under $Z \to -Z$, if $f(Z)$ is one solution, the other must
be $f(-Z)$.\footnote{Note that $f(Z)$ and $f(-Z)$ are not linearly independent solutions when $m^2=0$.
In this case, one must solve eq.\ \eref{hyper} for a second real solution.}
The general solution is therefore of the form
\be
F(Z) = a \, f(Z) + b \, f(-Z).
\ee
The hypergeometric function $f(Z)$ has a singularity at $Z = 1$, which translates to $d(x,y) = 0$.
In other words, the first term produces the well-known singularity along the lightcone.
The second term $f(-Z)$ is singular when $Z = -1$; that is, there is a singularity when points are
antipodal to each other.
The latter singularity is rather unconventional since antipodal points in de Sitter space are separated
by a cosmological horizon.
The Euclidean vacuum \cite{ghds, bd}, which we label by $\alpha = 0$, corresponds to setting $b=0$,
thus removing the antipodal singularity and achieving a smooth Minkowski limit.
It is also called the Euclidean vacuum, since it is selected by requiring a smooth Euclidean continuation,
or the adiabatic (thermal) vacuum, since it ensures that the behavior of the detector response function of
an Unruh detector is consistent with the principle of detailed balance \cite{burges}.


\section{The Operator Formalism: $\alpha$-vacua and $\alpha$-states}
\label{sec:opal}

Consider the mode expansion for the free field in the Euclidean vacuum:
\be
\phi(x) = \sum_n \left( a_n^E \phi_n^E(x) + a_n^{E\dagger} \phi_n^{E*}(x) \right).
\label{mode}
\ee
The vacuum $\ket{\al=0}$ is annihilated by the operators $a_n^E$ associated with the Euclidean mode basis
$\phi_n^E(x)$.
The modes can be chosen such that $\phi_n^E(x_A) = \phi_n^{E*}(x)$ \cite{allen}.

We can then define a new mode basis and associated annihilation operators
\begin{eqnarray} \label{modes}
\phi_n^\alpha(x) &=& \cosh\alpha_n\ \phi_n^E(x) + e^{-i\beta_n}\sinh\alpha_n\ \phi_n^{E*}(x), \\
a_n^\alpha &=& \cosh\alpha_n\ a_n^E - e^{i\beta_n} \sinh\alpha_n\ a_n^{E\dagger},
\label{squeezed}
\end{eqnarray}
with real $\alpha_n\in [0,\infty)$, $\beta_n\in [0,2\pi)$.
The vacuum $\ket{\alpha}$ arises as a Bogoliubov transformation of the Euclidean vacuum with mode and 
frequency independent coefficients $\alpha$, $\beta$.
It is annihilated by $a_n^\al$.
As time-reversal invariance requires that $\beta = 0$ \cite{allen}, we will henceforth drop this parameter.

Formally, we can write the $\alpha$-vacua as squeezed states of the Euclidean
vacuum \cite{qo}:
\be\label{formal1} 
a_n^\alpha = \CU a_n^E \CU^\dagger, 
~~~~~~ \ket{\alpha} = \CU \ket{0}, 
~~~~~~ \CU = A \exp\left[\sum_n \frac{1}{2} \left(C_n\,(a_n^{E\dagger})^2 - C_n^*\,(a_n^E)^2 \right) \right],
\ee
where $C_n = \al_n = \alpha = {\rm constant} \in \BR$.
Clearly, $a_n^E\ket{0} = 0$ implies that $a_n^\alpha\ket{\alpha} = 0$.
Although in eq.\ \eref{formal1}, $\CU$ appears to implement a unitary rotation of $\ket{0}$ to give the
vacuum state $\ket\al$, this is merely a formal statement: $\ket{0}$ and $\ket\al$ are not in the same 
Fock space, and $\alpha$ is a superselection parameter.\footnote{From the perspective of the conjectured
duality between a quantum theory of gravity in de Sitter space and a Euclidean conformal field theory
(CFT) on the boundary $\CI^\pm$ \cite{dscft, bdm1, bdm2}, different values of $\al$ arise as marginal 
deformations of the CFT \cite{bms}.}

The operators
\be
J_+ = \frac{1}{2}(a_n^E)^2, ~~~~~~ 
J_- = \frac{1}{2}(a_n^{E\dagger})^2, ~~~~~~ 
J_3 = \frac{1}{2}\{a_n^E, a_n^{E\dagger}\}
\ee
determine an $sl_2$ algebra.
The overlap between the Euclidean vacuum and the $\alpha$-vacuum is
\be
\langle 0 \ket\al = A \prod_n \cosh\alpha_n.
\ee
Setting $\bra{\alpha}\alpha\rangle = 1$ and performing standard manipulations using a matrix
representation of the $sl_2$ algebra, one finds that
\be
1 = \bra{\alpha}\alpha\rangle = 
|A|^2 \left(\prod_{n} \cosh 2\alpha_n \right). 
\label{norm}
\ee
As there are an infinite number of modes, $|A|^2$ must vanish.
Because the state $\ket\alpha$ cannot be normalized, it is not an excitation in the Fock space
constructed over $\ket{0}$.
Each of the $\alpha$-vacua defines the de Sitter invariant ground state of a different Hilbert space.

Choosing $\alpha_n$ mode-by-mode, we can, however, construct a normalizable state in the same Hilbert
space as $\ket{0}$.
For small $\alpha$, expanding the previous expressions requires that 
\be
0 \le |A|^2 = \left( 1 - 2 \sum_n \alpha(n)^2 + \ldots \right) \le 1.
\ee
More generally, if we are in $d$ dimensions, we expect the momentum density to scale as $k^{d-2}$.
Normalizability demands 
\be
\int dk\, k^{d-2} \log\cosh \alpha(k) < \infty,
\ee
which requires $\alpha(k)$ to decay faster than $k^{(1-d)/2}$.
By choosing an $\alpha(k)$ that exhibits a sufficiently fast falloff, we can construct states with
behavior characteristic of an $\alpha$-vacuum at low mode numbers but exponentially close to being like
the Euclidean vacuum at high mode numbers.
(Of course, such states will still not be de Sitter invariant.)
A state with fixed $\alpha$ up to some cutoff would naturally fall within this setting.
From the point of view of effective field theory, where we expect new physics to emerge beyond some
cutoff, this is perhaps a compelling realization of the ``$\al$-vacuum'' as an Euclidean excitation
\cite{glel}.
It is unreasonable to insist upon ``alphaness'' of the state at arbitrarily high mode number.

As well, it is tempting to think about generic initial conditions for inflation in this way.
Inflation can begin and end in an excited state, which will be some unspecified excitation over the
Euclidean ground state.\footnote{The Hartle-Hawking no boundary prescription for the wavefunction of the 
Universe \cite{HH} selects the Euclidean vacuum \cite{Laflamme}.}
\footnote{From the point of view of the Wheeler-de Witt equation, which is used in the mini-superspace approximation 
to justify the Hartle-Hawking vacuum, the issue of superselection sectors in the macroscopic theory is a real
concern.  The Hamilton constraint can be solved in the case of de Sitter space for general $\alpha$.}
Such an excitation will have certain modes populated.
That it is in the same Hilbert space as $\ket{0}$ ensures that the profile $\al(k)$ falls off
sufficiently fast at large momenta.
Inflation will take care of the rest.
Low momentum modes, whose character differs significantly from that of the Euclidean vacuum, are greatly
redshifted, so that the state today looks indistinguishable from the thermal vacuum.
Thus, it is conceivable that the Universe was $\alpha$-like at early times and is Euclidean at late times.
Such physics, we will argue, will be imprinted in fluctuations of the CMBR.

\section{The Functional Formalism}
\label{sec:funal}

\subsection{The wavefunctional of the $\al$-vacuum}

To further explore the physics of an $\al$-state, we find it useful to work in the functional formalism.
Following Refs.\ \cite{burges, Laflamme, jackiw} consider a scalar field in $\ds{d}$ with stress-energy
$T_{\mu\nu} = \partial_\mu \phi \partial_\nu \phi - \eta_{\mu\nu} {\cal{L}}$.
The de Sitter symmetries (boosts and rotations) are generated by
\bea
M_{0i} &=& \int d^dx\, \sqrt{-g}\, (x^i T^{00} - tT^{0i}) = \int d^dx\, \sqrt{-g}\, \frac{1}{2} x^i(\pi^2 +(\nabla \phi)^2), \\
M_{ij} &=& \int d^dx\, \sqrt{-g}\, (x^i T^{0j} - x^jT^{0i}) = \int d^dx\, \sqrt{-g}\, \phi (x^i \partial_j - x^j\partial_i) \pi.
\eea
The scalar field $\phi(x)$ and its conjugate momentum 
$\pi(x) \equiv -i\frac{1}{\sqrt{-g(x)}}\,\frac{\delta}{\delta\phi(x)}$
satisfy canonical equal-time commutation relations 
$[\phi(x), \phi(x')] = [\pi(x), \pi(x')] = 0$ and
$[\phi(x), \pi(x')] = \frac{i\delta^{(d-1)}(\x-\x')}{\sqrt{-g(x')}}$.
To be de Sitter invariant, the wavefunctional of the vacuum must be annihilated by the symmetry generators.

Following \cite{burges}, consider a gaussian ansatz for the vacuum wavefunctional:
\begin{equation}
\langle \phi|0 \rangle  = \exp\left(-\frac{1}{2} \int d^dx\, \sqrt{-g(x)}\, \int d^dx'\, \sqrt{-g(x')}\, \phi(x) F(x,x') \phi(x')\right).
\end{equation}
By imposing $M \bra{\phi}0\rangle = 0$, one finds from rotational symmetry that $F(x,x') = F(|x-x'|)$.
This means that $F(x,x')$ can be expanded in terms of $(d-1)$-dimensional spherical harmonics:
\begin{equation}
F(|x-x'|) = \sum_{\ell,\vec{m}} f_\ell \, Y^{*}_{\ell\vec{m}} \,  Y_{\ell\vec{m}}.
\end{equation}
The boost condition for a fixed $x^i$ determines the form of $f_\ell$ up to a one complex parameter
ambiguity \cite{burges, jackiw} corresponding to the choice of $\alpha$.
(The explicit formula for $f_\ell$ is not particularly illuminating and can be found in Ref.\ \cite{burges}
for $d=2,3,4$.)
It can be shown that the Hartle-Hawking Euclidean prescription for the evaluation of the de Sitter
wavefunctional restricts us to the adiabatic vacuum \cite{HH, Laflamme}.

The general wavefunctional (for arbitrary $\alpha$ parameter) is given by \cite{burges}
\be \label{wf}
\Phi_{\alpha}(\phi)= e^{i \alpha \hatS(\phi)} \Psi_{\alpha=0}(\phi),
\ee
where $\hatS$ provides the action of the antipodal map
\be
[i\hatS, \phi(x)] = \frac12 \phi(x_A).
\ee
The antipodal map $\hatS$ is written as a spatial integral:
\be
\hatS = \frac12 \int_{\Sigma_{t=0}} d^{d-1}x\, \sqrt{\gamma(x)}\, \phi(x_A) \pi(x).
\label{antip}
\ee
The integration is in global coordinates over the $\Sigma_{t=0}$ surface.
This is the unique spatial section which contains both a point and its antipode.
$\gamma(x)$ is the pull-back of the de Sitter metric onto the $t=0$ surface.
To simplify our expressions, we shall suppress from now on factors of the determinant of the (pull-back)
metric in our discussion of the scalar field.
(We shall always treat the momentum as a scalar, rather than as a density, however.)
In principle, a similar map $\hatS$ will exist for each field in the theory.
We shall return to this point later.

Setting $\alpha = 0$ in eq.\ \eref{wf} corresponds to selecting the Euclidean vacuum wavefunctional
$\Psi$, which satisfies the functional Schr\"odinger equation
\be
\frac{1}{2} \int d^{d-1}x\, 
\left( \pi(x)^2 + (\nabla\phi)^2 + \meff^2 \phi(x)^2 \right) \Psi(\phi) = 
\omega \Psi(\phi).
\ee
We write the effective mass $\meff^2 := m^2 + \xi R$ to include any quadratic coupling of $\phi(x)$ to
the curvature.
For non-zero, constant $\alpha$, the Schr\"odinger equation is 
\be
e^{i\alpha \hatS(\phi)} \cdot
\frac12 \int d^{d-1}x\,\left( \pi(x)^2 + (\nabla \phi)^2 + \meff^2 \phi(x)^2 \right) 
\cdot e^{-i\alpha \hatS(\phi)} \Phi(\phi) = \omega \Phi(\phi).
\label{scheq}
\ee
We will consider only the Hamiltonian of a free scalar field in de Sitter space.\footnote{For an analysis
of interacting scalar fields in de Sitter space and the implications for cosmology, see Ref.\ \cite{malda}.} 

\paragraph{Comparison to QCD $\theta$-vacua} ${}$ \\
The functional description of the $\al$-vacua is {\em formally} similar to that of $\theta$-vacua in
QCD \cite{theta}.  
Since we shall later consider promoting the parameter $\al$ to a dynamical field, it is useful to make
a brief comparison with QCD.

The general gauge invariant wavefunctional for a $\theta$-vacuum is
\be
\Phi_{\theta}(\vec{A})= e^{i \theta W(\vec{A})} \Psi_{\theta = 0}(\vec{A}),
\ee
where by {\em vacuum}, we mean a state that solves the appropriate Gauss constraint.
$W(\vec{A})$ is the Chern-Simons form, which under large gauge transformations changes by an integer.
The variation of $W(\vec{A})$ with respect to the gauge connection $\vec{A}$ gives the Yang-Mills magnetic field
$\vec{B}_a = \nabla \times \vec{A}_a - \frac{1}{2} g f_{abc} \vec{A}_b \times \vec{A}_c$,
where $g$ is the Yang-Mills gauge coupling and $f_{abc}$ are the structure constants of the gauge group;
$\frac{\delta W}{\delta {\vec{A}}} = \frac{g^2}{8\pi^2} \vec{B}$.
$\Psi(\theta = 0)$ satisfies a functional Schr\"odinger equation: 
\be
\int d^3x\,\frac{1}{2}\left( - \frac{\delta^2}{\delta \vec{A}_a^2} +\vec{B}_a^2 \right) \Psi (\vec{A}) = \omega \Psi (\vec{A}).
\ee
The Hamiltonian density is the usual $\frac{1}{2} (\vec{E}^2 + \vec{B}^2)$,
and $\vec{E}$ is the canonical momentum $-\dot{\vec{A}}$.
The Schr\"odinger equation for $\Phi$ for a non-zero $\theta$
(taking into account the variation of $W(\vec{A})$) reads
\be
\int d^3x\,\frac{1}{2}\left( \left(-i \frac{\delta}{\delta \vec{A}_a} - \frac{\theta g^2}{8\pi^2}\vec{B}_a \right)^2
+ \vec{B}_a^2 \right) \Phi(\vec{A}) = \omega \Phi(\vec{A}).
\ee
By doing the phase space path integral 
\be
\int D\vec{E}_a\,D\vec{A}_a\,
\exp\left(-i \int d^3x\,\left[\vec{E}_a \cdot \dot{\vec{A}}_a + 
\frac{1}{2}\left(\vec{E}_a + \frac{\theta g^2}{8\pi^2}\vec{B}_a\right)^2 +
\frac{1}{2} \vec{B}_a^2\right]\right),
\ee
one finds, upon integrating over $\vec{E}_a$, the expected term
$\theta \dot{\vec{A}} \cdot \vec{B} = \theta \vec{E} \cdot \vec{B}$ 
in the effective Lagrangian \cite{theta}.
In covariant notation, this is $\theta F \wedge \widetilde{F}$.
The parameter $\theta$ can be promoted into a field (the axion), by adding
a canonical kinetic term to the Lagrangian \cite{theta}.
The new operator $\theta F \wedge \widetilde{F}$ is local, dimension five, and CP violating. 
This interaction is suppressed by some high-energy scale accommodating experimental bounds on CP violation.
(Various phenomenological implications have been reviewed in Ref.\ \cite{axion}.)

Although we may draw inspiration from this example, the two equations 
\bea
&& \Phi_{\theta}(\vec{A}) = e^{i \theta W(\vec{A})} \Psi_{\theta = 0}(\vec{A}), \\
&& \Phi_{\alpha}(\phi)= e^{i \alpha \hatS(\phi)} \Psi_{\alpha=0}(\phi) 
\eea
exist on separate and unequal footings.
The logic of $\Phi_\theta(\vec{A})$ is that it is the most general wavefunctional
consistent with gauge invariance given that we are working with projective
representations of the Hilbert space of Yang-Mills theory.
The logic of $\Phi_\alpha(\phi)$ is that it is the most general wavefunctional
consistent with de Sitter invariance.
Gauge invariance is a {\em local} symmetry, whereas de Sitter 
invariance is a {\em global} statement about the geometry.
(However, the large gauge transformations under which the Yang-Mills vacuum
wavefunctional is invariant, are an analogue of the de Sitter isometry,
which is, in a sense, the set of large diffeomorphisms.)
As well, while $W(\vec{A})$ is a {\em functional} of the gauge connection that
rotates $\Psi_{\theta=0}(\vec{A})$ by a phase, $\hatS(\phi)$ is an {\em operator}
that encodes a non-local interaction between $\pi(x)$ and $\phi(x_A)$.
Indeed, eq.\ \eref{wf} is yet another description of the formal
unitary transformation of vacua that we have encountered previously.
We also note that the short distance structure of the singularities is
different in each of the $\alpha$-vacua, whereas
no analogous statement exists for the $\theta$-vacua.
While significant similarities do exist between the wavefunctional approaches
to the $\theta$-vacua of QCD and the $\alpha$-vacua of de Sitter space,
there are crucial differences, and we should proceed with care in exploiting this analogy.

\paragraph{Field theory in an $\al$-vacuum} ${}$ \\
To consider fields in an $\al$-vacuum, we compute the left hand side of 
the functional Schr\"odinger equation \eref{scheq}. 
Taylor expanding the exponentials and iterating commutators with $i\alpha \hatS$,
we find 
\be
\tpi(x) := e^{i\alpha \hatS(\phi)} \cdot \pi(x) \cdot e^{-i\alpha \hatS(\phi)} =
\pi(x) - \frac{\alpha}{2} \pi(x_A) + \frac{\alpha^2}{8} \pi(x) + \ldots,
\ee
which can be resummed to give
\be \label{j1}
\tpi(x) = \cosh\frac{\alpha}{2}\ \pi(x) - \sinh\frac{\alpha}{2}\ \pi(x_A);
\ee
and similarly,
\be \label{j2}
\tphi(x) := e^{i\alpha \hatS(\phi)} \cdot \phi(x) \cdot e^{-i\alpha \hatS(\phi)} = 
\cosh\frac{\alpha}{2}\ \phi(x) + \sinh \frac{\alpha}{2}\ \phi(x_A).
\ee
The gradient $\nabla\phi$ does not affect the analysis in a crucial way.
In particular, we find 
\be
e^{i\alpha \hatS(\phi)} \cdot \nabla_x \phi(x) \cdot e^{-i\alpha \hatS(\phi)} 
= \nabla_x \tphi(x).
\ee
The field redefinitions act like a canonical transformation.
They preserve the commutation relations
\be
[\phi(x),\pi(x')] = [\tphi(x),\tpi(x')] = i\delta^{(d-1)}(x-x').
\ee
Conjugation by the antipodal map $\hatS$ implements the Bogoliubov
transformation on the fields!
Note that it is only for real $\alpha$ that the new Hamiltonian we obtain
in this way is Hermitian.

Eqs.\ \eref{j1} and \eref{j2} imply that if $\alpha$ is a non-zero, real
constant, then $\tphi$ is in the $\alpha$-vacuum if the original field
$\phi$ is in the Euclidean vacuum.
The two-point Wightman function for $\tphi$ is precisely as expected:
\bea
G_{\alpha}(x,y) &:=& \bra{\alpha}\tphi(x) \tphi(y)\ket{\alpha} \nonumber \\ 
&=& \cosh^2\alpha\ G_E(x,y) + \sinh^2\alpha\ G_E(x_A, y_A) +
\frac{1}{2}\sinh 2\alpha\ [G_E(x_A, y) + G_E(x,y_A)].
\label{twopt}
\eea

Equations of motion for $\tphi(x)$ obtain from the Lagrangian 
\be
L = \int d^{d-1}x\, \tpi(x) \dot{\tphi}(x) - \Ham
\ee 
with $\Ham$ the Hamiltonian
\be
\Ham = \frac{1}{2} \int d^{d-1}x\, \left( \tpi(x)^2 + (\nabla \tphi)^2 + \meff^2 \tphi(x)^2 \right)
\label{hamiltonian}
\ee
defined using the relations \eref{j1} and \eref{j2}.
The action written with respect to the original variables $\phi(x)$ and
$\pi(x)$ is clearly non-local, but in terms of the new variables $\tphi(x)$
and $\tpi(x)$ it is perfectly local!
Conjugation by the antipodal map converts a {\em non-local} description
involving fields in the Fock space over the Euclidean vacuum to a {\em local}
description involving fields in the Fock space over the $\alpha$-vacuum.
This is as we expect.
As $\ket0$ and $\ket\al$ are the ground states of different Hilbert
spaces, $\phi(x)$ and $\tphi(x)$ are local excitations of different vacua.
The structure of the vacuum determines the most natural field variable for
writing the Lagrangian.
Regarding {\em either} the Hartle-Hawking vacuum of eq.\ \eref{hamiltonian},
with $\tphi(x)$ as the fundamental field in the $\alpha$-vacuum, or 
equivalently the Hartle-Hawking vacuum with $\phi(x)$ as the fundamental  
field, lead to the same $\alpha$-states.

\subsection{The wavefunctional of the $\al$-state}

We shall now adapt the functional formalism to the study of $\al$-states.
We want to write the analogue of the Schr\"odinger equation \eref{scheq} for
the wavefunctional $\langle\phi\ket\al$, where now $\ket\al$ is a normalizable
excitation over the Euclidean vacuum $\ket{0}$.

It is instructive to recall that an $\al$-state in the operator formalism
corresponds to the profile $\al(k)$.
This is an expression in {\em momentum} space.
The function $\al(k) = {\rm constant}$ defines the $\al$-vacuum.
The antipodal map $\hatS$ in eq.\ \eref{antip}, by contrast, is defined as
an integral over the $\Sigma_{t=0}$ spatial surface.
In the expression
\be
\Phi_{\alpha}(\phi)= e^{i \alpha \hatS(\phi)} \Psi_{\alpha=0}(\phi)
\ee
for the de Sitter vacuum wavefunctional, $\al$ simply appears as a coefficient
in the exponential.
We wish to promote this coefficient to a function of position consistent
with the intuition that {\em constant} $\al$ in {\em momentum} space
corresponds to a {\em $\delta$-function} distribution of weight $\al$
in {\em position} space.

We make the ansatz
\be
\Sal = \lambda \int d^{d-1}x\, d^{d-1}y\, \al(x-y) \phi(x_A) \pi(y). \label{Sal}
\ee
Note that $\lambda$ may be dimensionful: $\dim\ \lambda + \dim\ \alpha = d-1$.
Clearly, putting $\al(x-y) = \al\, \delta^{(d-1)}(x-y)$ recovers the
antipodal map $\hatS$ from before.

We note that $\al$ appears in the antipodal map \eref{Sal} as a function of
the {\em difference} in position of two points on the $\Sigma_{t=0}$ spatial
surface integrated over the surface.
Taking $\al(x-y) = \al(|x-y|)$ is consistent with physics in the $\al$-state
being homogeneous and isotropic.
If we regard the scalar as being the inflaton field, this is a desirable
feature of the cosmology.
In global coordinates in $d=4$, homogeneity and isotropy imply that $\alpha(x-y)$ is
really $\alpha(\tr(g^{-1} h))$, where $g$ and $h$ are two points on $S^3$ 
via the identification of $S^3$ with $SU(2)$, so that $g$ and $h$ are $SU(2)$
matrices. This follows, for example, from the fact that $d(g,h)$, the distance on $S^3$,
obeys $2\cos d(g,h) = \tr(g^{-1} h)$.
In inflationary coordinates, $\alpha(x-y)$ denotes a function of the 
difference between $x$ and $y$. The map between these two definitions of 
$\alpha$ is complicated; they live on different equal time surfaces.

The Fourier transform\footnote{Strictly speaking, in global coordinates we 
should perform a discrete Fourier transform and write the result as a sum
over momentum modes. For ease of notation, we prefer to write integral
expressions instead.} of eq.\ \eref{Sal} is
\be
\Sal = \lambda \int \frac{d^{d-1}k}{(2\pi)^{d-1}}\, \al(k) \phi(k) \pi(k).
\ee
(In writing this expression, we have employed the fact that de Sitter space is
a maximally symmetric space.\footnote{To be explicit, the calculation proceeds as below:
\bea
\Sal &=& \lam \int d^{d-1}x\, d^{d-1}y\, \frac{d^{d-1}k}{(2\pi)^{d-1}}\, \al(k) e^{-ik\cdot(x-y)} \phi(x_A) \pi(y) \nn \\
&=& \lam \int \frac{d^{d-1}k}{(2\pi)^{d-1}}\, \al(k) \cdot \int d^{d-1}x\, e^{-ik\cdot x} \phi(x_A) \cdot \int d^{d-1}y\, e^{+ik\cdot y} \pi(y) \nn \\
&=& \lam \int \frac{d^{d-1}k}{(2\pi)^{d-1}}\, \al(k) \cdot \int d^{d-1}x\, e^{-ik\cdot x_A} \phi(x) \cdot \pi(k) \nn \\
&=& \lam \int \frac{d^{d-1}k}{(2\pi)^{d-1}}\, \al(k) \phi(k) \pi(k). \nn
\eea
Integrating the pair $(x,x_A(x))$ over the spatial slice is the same as integrating the pair $(x_A(x),x)$ over the slice.
More formally, one writes
$$
1 = \int d^{d-1}z\, \delta^{(d-1)}(z-x) = \int d^{d-1}z\, \delta^{(d-1)}(z_A-x),
$$
and performs the $x$ integration with the $\delta$-function and relabels variables at the end.
In the last step we have used the Hermiticity properties of the spherical harmonics,
$Y_{l,\vec{m}}(x_A) = Y_{l,\vec{m}}(x)^*$.})
We note that the terms in the integrand are in accord with the expectation of
the operator formalism:
\be
\phi(k)\pi(k) \Longrightarrow 
(a_k^\dagger + a_k) (a_k^\dagger - a_k) \sim \left((a_k^\dagger)^2 - (a_k)^2\right).
\label{ops}
\ee
The right hand side of eq.\ \eref{ops}, we recall, appears in the definition
of $\CU$ in the formal expression for the squeezed state $\ket\al =
\CU \ket{0}$ (\cf eq.\ \eref{formal1}).

To write the Schr\"odinger equation corresponding to a wavefunctional in an
$\al$-state, the analysis proceeds exactly as above.
We simply conjugate the free field Hamiltonian by $e^{i\Sal}$.
We find that the transformed Hamiltonian is of the same form as the original
Hamiltonian, but is written in terms of the new field variables
\bea
\tphi(x) &:=& e^{i\Sal} \cdot \phi(x) \cdot e^{-i\Sal} \nn \\
&=& \phi(x) + \lambda \int d^{d-1}y\, \al(y-x) \phi(y_A) +
\frac12 \lambda^2 \int d^{d-1}y\, d^{d-1}z\, \al(y-x) \al(z-y_A) \phi(z_A) + \ldots, \nn \\ 
\label{tphi} \\
\tpi(x) &:=& e^{i\Sal} \cdot \pi(x) \cdot e^{-i\Sal} \nn \\
&=& \pi(x) - \lambda \int d^{d-1}y\, \al(x_A-y) \pi(y) +
\frac12 \lambda^2 \int d^{d-1}y\, d^{d-1}z\, \al(x_A-y) \al(y_A-z) \pi(z) + \ldots. \nn \\
\label{tpi}
\eea
The canonical equal-time commutation relation
\be
[\phi(x),\pi(x')] = [\tphi(x),\tpi(x')] = i\delta^{(d-1)}(x-x')
\ee
is preserved by this transformation.\footnote{The algebra is as follows:
\bea
[\tphi(x),\tpi(x')] &=&  [\phi(x),\pi(x')] + \lam\, \al(x'_A-x) - \lam\, \al(x'_A-x)
+ \frac12 \lam^2 \int d^{d-1}y\, \al(y-x) \al(x'_A-y_A) \nn \\
&&
+ \frac12 \lam^2 \int d^{d-1}y\, \al(x'_A-y) \al(y_A-x)
- \lam^2 \int d^{d-1}y\, \al(y-x) \al(x'_A-y_A)
+ \ldots. \nn
\eea
The cancellation of terms at $\CO(\lambda^2)$ demands that we make use
of the invariance of the integrand under $y\leftrightarrow y_A$, when $y$
is the variable of integration.
This once again invokes the maximal symmetry of de Sitter space.}

In the case where $\al(x-y)$ is nearly a $\delta$-function, the field
redefinition of eq.\ \eref{tphi} presents a smearing of eq.\ \eref{j2}.
The even powers of $\lambda$ in the expansion of $\tphi(x)$ coarsen out
$\phi$ near $x$, while the odd powers coarsen out $\phi$ near $x_A$.
This is, however, not a case of there being an extra antipodal source in
an interaction Lagrangian, as has been proposed for an $\al$-vacuum \cite{collins}.
As we are working in the Euclidean vacuum, the singularity structure of the
Green function is unchanged.
We need not adopt a non-standard Feynman prescription in this theory.

We can equally write momentum space expressions $\tphi(k)$ and $\tpi(k)$.
The advantage of this is that we can sum the series to get closed form expressions:
\bea
\tphi(k) &=& e^{i\Sal} \cdot \phi(k) \cdot e^{-i\Sal} \nn \\
&=& \phi(k) + \lam\, \al(-k) \phi(-k) + \frac12 \lam^2\, \al(-k)\al(k)\phi(k) + \ldots \nn \\
&=& \cosh s(k)\, \phi(k) + \lam\, \al(-k) \frac{\sinh s(k)}{s(k)}\, \phi(-k), \\
\tpi(k) &=& e^{i\Sal} \cdot \pi(k) \cdot e^{-i\Sal} \nn \\
&=& \pi(k) - \lam\, \al(-k) \pi(-k) + \frac12 \lam^2\, \al(-k)\al(k)\pi(k) + \ldots \nn \\
&=& \cosh s(k)\, \pi(k) - \lam\, \al(-k) \frac{\sinh s(k)}{s(k)}\, \pi(-k), 
\eea
where $s(k) := \lam \sqrt{\al(k)\al(-k)}$.
Fourier transformation recovers the position space expressions we have seen previously.

The Hamiltonian density, upon non-local field redefinition, is simply
that of a free field:
\be
\widehat\CH = \frac{1}{2} \left( \tpi(x)^2 + (\nabla \tphi)^2 + \meff^2 \tphi(x)^2 \right).
\label{ch}
\ee
Whereas the field $\tphi(x)$ is the appropriate local variable to work with
when in an $\al$-vacuum, it is not clear here {\em a priori} whether we should
work with $\phi(x)$, which is an excitation of the Euclidean vacuum $\ket0$,
or $\tphi(x)$, which is an excitation of the state $\ket\al$, itself a
squeezed excitation over $\ket0$.
The Hamiltonian density expressed in terms of $\phi(x)$ and $\pi(x)$ is
clearly non-local in that it involves integrations over the entire
$\Sigma_{t=0}$ surface.
We interpret the observation that the Hamiltonian density, when written
in terms of the new field variables $\tphi(x)$ and $\tpi(x)$, looks local
while secretly being non-local in terms of the old field variables $\phi(x)$
and $\pi(x)$ as a signal that the physics of an $\al$-state is nothing but
the highly entangled physics of scalar excitations over the Euclidean 
ground state.
We are simply working with a state in which there are long-range (Hubble
scale) correlations.
There is no experiment that demonstrates that field theory exhibits the usual decoupling 
at the scale of the cosmological horizon.
Local measurements by a de Sitter observer are completely consistent with
this philosophy.
Indeed, a signal sent from the antipode at global time $t=-\infty$ (on $\CI^-$)
only reaches us at time $t=+\infty$ (on $\CI^+$).
Antipodal correlations do not ring the death knell for causality.
As we have performed a canonical transformation via a unitary
rotation, there can be no inconsistency in the resulting physics.
We merely choose to work in an entangled state where the structure of the
theory looks different.
Such antipodal entanglements on the $\Sigma_{t=0}$ surface arise from some
correlated initial conditions at early times.
If we go beyond free field theory to include interactions, the 
perturbative consistency of the theory in the Euclidean vacuum ensures the perturbative consistency
of the theory in the $\al$-state.\footnote{Coupling matter to the metric
beyond the free field level will generically stretch the Penrose diagram
giving a tall de Sitter space \cite{rm}.
In such a geometry, a point and its antipode are in causal contact before
the end of time.
One might na\"{\i}vely conclude that antipodal non-localities will lead in
such a setting to acausal or non-unitary physics.
This is not the case.
The initial conditions are responsible for an EPR-like entanglement.}

\section{Physics of $\al$-states}
\label{sec:aldyn}

\subsection{$\al$ as a fixed profile}

Let us suppose that $\al(k)$ is a given function that defines an $\al$-state.
The Hamilton equations of motion for $\tphi$ in this $\al$-state on the
$\Sigma_{t=0}$ surface simply give the Klein-Gordon equation:
\be
0 = (\Box_x - \meff^2) \tphi(x),
\label{kg}
\ee
where $\Box_x$ is the usual Laplace-Beltrami operator in curvilinear
coordinates.
We therefore expect that the functional form of $\tphi(x)$ is the same as
the functional form of $\phi(x)$ in the Euclidean vacuum.
The only difference is that the mode expansion
\be
\tphi(x) = \sum_n \left( \ta_n \tphi_n(x) + \ta_n^\dagger \tphi_n^*(x) \right)
\ee
is written in terms of operators that act on the $\al$-state:
$\ta_n \ket\al = 0$.
These operators satisfy the Heisenberg algebra $[\ta_m, \ta_n^\dagger] =
\delta_{mn}$.
The $\tphi_n(x)$ are linear combinations of the $\phi_n^E(x)$ mode solutions
\eref{mode} as they solve the same wave equation.

Written in terms of the field $\phi(x)$, the Klein-Gordon equation is
highly non-trivial:
\be
0 = (\Box_x - \meff^2) \left( \phi(x) + \lam \int d^{d-1}y\, \al(y-x) \phi(y_A) + 
\frac12 \lambda^2 \int d^{d-1}y\, d^{d-1}z\, \al(y-x) \al(z-y_A) \phi(z_A) + \CO(\lam^3) \right)
\label{kgnl}
\ee
Note that $\Box_x$ acts on $\al(y-x)$ under the integrals.
Eq.\ \eref{kgnl} is a non-local integro-differential equation of motion.
It is non-local in the sense that it involves an integration over the entire
$\Sigma_{t=0}$ surface.\footnote{
Such equations, we note, have appeared in the literature in other contexts,
\eg stress-strain relations \cite{rayleigh}, radiative transfer
\cite{chandra}, and neurophysiology \cite{neuro}.
Wiener-Hopf techniques sometimes permit these equations to be solved for
certain kernels when the integral is one-dimensional ($d=2$) \cite{baraff}.}
Eq.\ \eref{kg} dumps the non-local structure of the Klein-Gordon equations
into the action of the $\ta_n$ on the $\al$-state. 

So far, we have worked exclusively on the $\Sig$ surface.
How do we evolve the theory off the surface?
The usual Hamiltonian evolution
\be
\tphi(T,x) = e^{i\widehat{H}T} \tphi(x) e^{-i\widehat{H}T}
\ee
propagates the solution from the $\Sig$ spatial surface forward in time to the
$\Sigma_{t=T}$ spatial surface.
For any $t\ne 0$, the correlations in the expansion for $\tphi(x)$ lie 
along a single spatial section.
One would na\"{\i}vely conclude that
the antipodal correlation induced by the $\al(x-y)$ is {\em not} an antipodal
interaction in the de Sitter spacetime.  
Rather the correlations are antipodal on the $S^{d-1}$ sphere at time $T$. 
This is inconsistent with the expectation for physics in an $\al$-vacuum.
When $\al(x-y) = \al\, \delta^{(d-1)}(x-y)$, we expect to recover the singular
behavior of the Green function both along the light cone {\em and} at the
antipode in the de Sitter spacetime.

Recall that we can write the mode expansion for $\phi(x)$ in the Euclidean vacuum as
\bea
&& \phi(x) = \sum_n ( a_n^E \phi_n^E(x) + a_n^{E \dagger} \phi_n^{E*}(x) ) 
= \sum_n ( a_n^\alpha \phi_n^\alpha(x) + a_n^{\alpha\dagger} \phi_n^{\alpha*}(x) ),
\eea
where the operators and mode functions are as in eqs.\ \eref{modes}--\eref{squeezed}.
The scalar field in an $\al$-vacuum is
\bea
&& \phi^\alpha(x) = \cosh\alpha\, \phi(x) + \sinh\alpha\, \phi(x_A)
= \sum_n ( a_n^E \phi_n^\alpha(x) + a_n^{E \dagger} \phi_n^{\alpha*}(x) ).
\eea
The Euclidean mode basis can be chosen \cite{allen} such that $\phi_n^*(x) = \phi_n(x_A)$.
Thus, if one considers a field prepared in the Euclidean-vacuum on the $\Sig$ surface and
implements the usual Hamiltonian evolution, $\phi(x,T) = e^{i\Ham T} \phi(x,0) e^{-i\Ham T}$,
the field knows about the backward time evolution as well because the mode functions
are chosen to be antipodally related.
The field $\phi^\alpha(x)$ in the $\al$-vacuum and the field $\tphi(x)$ in the $\al$-state
inherit this property.\footnote{We are working in the Heisenberg picture, where operators
evolve. In the Schr\"odinger picture, the state $\ket\al$ evolves non-trivially, even for an
$\al$-vacuum. Translation in global time is not an isometry of de Sitter space.}
The latter result, that $\tphi^*(T,x) = \tphi(-T,x_A)$, where $x_A$ is the 
antipode on the spherical section, relies on the homogeneity and isotropy of the kernel
$\al(x-y)$ in $\Sal$.

\paragraph{A second look at time evolution} ${}$ \\ 
One can also apply the intuition of the eternal anti-de Sitter black hole to the problem
of time evolution \cite{juan}.
This alternate perspective is completely consistent with the prior reasoning in terms
of an explicit mode analysis.
The state $\ket\tphi$ is doubled, as in thermofield dynamics.
To every operator $\CO$, we associate its conjugate $\overline\CO$.
(The algebraic rules for conjugation are given in Ref.\ \cite{Umezawa}, for example.) 
In particular, the free Hamiltonian, which is the generator of time translation, is
\be
\widehat{H}_{TFD} = {H} - {\overline{H}}
\ee
and commutes with the squeezing operator
\be
\CU_{TFD} = A \exp\left[\sum_n \frac{1}{2} \left(C_n\, a_n^{\dagger} \bar{a}_n^{\dagger} - C_n^*\, a_n \bar{a}_n \right) \right].
\ee
The canonical equations of motion are
\bea
i\hbar\, \frac{\pa}{\pa t}\CO = [\CO, {H}], &&
i\hbar\, \frac{\pa}{\pa t}\overline\CO = -[\overline\CO, {\overline{H}}].
\eea
Hamiltonian evolution carries us forward and backward in time from the $\Sig$ surface:
\be
\ket{\tphi(T,x)}_{TFD} = e^{i{H} T} \ket{\tphi(0,x)} \otimes e^{-i{\overline{H}} T} \ket{\overline{\tphi(0,x)}}.
\ee
The first term in the tensor product propagates the field $\tphi(0,x)$ forward in time 
to the $\Sigma_{t=T}$ hypersurface.
The second term propagates the conjugate field $\overline{\tphi(0,x)}$ backward
in time to the $\Sigma_{t=-T}$ hypersurface.
In this way, antipodal correlations are maintained in the full de Sitter
spacetime.
As well, the boundary data on $\CI^-$ are correlated to the boundary data
on $\CI^+$.
Hence, if we accept the proposed de Sitter/CFT correspondence \cite{dscft}, the
Euclidean conformal field theories on the spheres at $t=\pm\infty$ are 
naturally entangled.
This is in the same spirit as the de Sitter holography discussed in Refs.\ \cite{bdm1, bdm2}.

The thermofield dynamics provides a compelling picture for the evolution of the state in
static coordinates where there exits a time-like Killing isometry. 
Consider the two static patches $N$ and $S$. 
The arrow of time will point in the opposite direction at the antipode, 
and these patches will cover the $\Sigma_{t=0}$ spatial surface.
A squeezed state with energy $\om$ is \cite{bdm2}
\be
\ket{\tphi} = C\exp\left[\frac{\cosh\al\, e^{-\pi\om} + \sinh\al}{\cosh\al\, + \sinh\al\, e^{-\pi\om}} a^\dagger_N a^\dagger_S \right] \ket{0}.
\ee
(The expression as written is for an $\al$-vacuum, but generalizes to an $\al$-state.)
The vacuum $\ket{0} = \ket{0}_N \otimes \ket{0}_S$.
The operators $a^\dagger_N$ and $a^\dagger_S$ act on different sectors of the product Hilbert space. 
Preparing this state is in perfect accord with thermofield dynamics where a squeezed state is defined 
through the action of $a^\dagger \bar{a}^\dagger$.
Time evolution on the static patches is just the usual Hamiltonian evolution.
The state evolves such that the field at $t=+T$ is correlated to the field at the antipode at $t=-T$,
as is required.

\paragraph{Phenomenology of $\al$-states} ${}$ \\
How can we tell whether we are living in an $\al$-state?
If we take the scalar field to be the inflaton, we can compute the power
spectrum for $\tphi(x)$, an excitation over $\ket\al$, and compare this 
to the power spectrum for $\phi(x)$, an excitation over $\ket0$.
We shall now switch to conformal coordinates where $\dS{d}$ is written
in terms of the embedding
\bea
&& X^d - X^0 = H^{-2} \tau^{-1}, \nn \\ 
&& X^i = H^{-1} \tau^{-1} x^i, \;\;\; i=1,2,\ldots,d-1, \\ 
&& X^d + X^0 = \tau - \tau^{-1} \sum_i (x^i)^2, \nn 
\eea
so that the de Sitter line element is
\be
ds^2 = H^{-2} \tau^{-2} \eta_{\mu\nu} dx^\mu dx^\nu.
\ee
The union of two coordinate patches, $\{\tau\in(-\infty, 0), \x\in\BR^{d-1}\}
\bigcup \{\tau\in(0, \infty), \x\in\BR^{d-1}\}$, covers $\dS{d}$ and the
antipodal point of $x=(\tau,\x)$ is $x_A=(-\tau,\x)$.  
In these coordinates $\tau=0$ corresponds to the boundaries $\CI^\pm$.
In defining the antipodal map, we have worked on the $t=0$ hypersurface in
global coordinates.
In conformal coordinates, the $S^{d-1}$ at $t=0$ is given by
\bea
&& \tau = H^{-1}\, 1/\omega^d = H^{-1}(\sin\theta_1 \ldots \sin\theta_{d-1})^{-1}, \\
&& x^i = H^{-1}\, \omega^i/\omega^d, \;\;\; i = 1, 2, \ldots, d-1,
\eea
where the $\theta_i$ are angle variables on the sphere that define the coordinates $\om^i$.
The antipodal transformation involves all $(\tau,x^i)$ consistent with this change of variables.
In particular, it includes an integral over points in {\em both} flat patches.
(Usually, one considers only a single coordinate patch and integrates (traces over) the other patch, 
or assigns vacuum expectation values to the degrees of freedom across the horizon and uses mean field theory.)

The calculation of the power spectrum relies on working in
inflationary coordinates.
We must choose an appropriate parametrization of the $d$-dimensional
de Sitter hyperboloid in $(d+1)$-dimensional Minkowski space.
In defining the antipodal transformation, we have made repeated use
of the maximal symmetry of the de Sitter geometry and the homogeneity
and isotropy of the argument of $\al$, which is a distance.
Although in global coordinates we work with $\alpha$ on the $\Sigma_{t=0}$
surface, we could also define the kernel on other Cauchy surfaces that are
preserved by the antipodal map.
The most natural extension to these surfaces is to define $\alpha$ in terms
of the embedding: $\al(x-y) = \alpha(|X(x)-X(y)|)$.
(In the inflationary coordinates, for $x=(\tau,x^i)$, $y=(\tau',y^i)$,
we will find that
\be
|X(x) - X(y)| = H^{-1} \left( \frac{-(\tau - \tau')^2 + \sum_i (x^i - y^i)^2}{\tau\tau'} \right)^{1/2},
\ee
so indeed the distance in the embedding space on an equal $\tau$ surface
is simply proportional to $|\vec{x}-\vec{y}|$.)
It will therefore make sense to calculate the momentum space
autocorrelation function of the scalar as an excitation in the
Euclidean vacuum through a Fourier transform using the previously
described conformal coordinates, and indeed, we will see that our
results match up with certain calculations in the inflationary
patch using other techniques.

We shall compute 
$\widetilde{P}(k) = \bra0 \left( |\tphi(k)|^2 \right) \ket0$
and compare this to 
$P(k) = \bra0 \left( |\phi(k)|^2 \right) \ket0$.
For definiteness, we choose to look at a massless, conformally coupled scalar
field in $d=4$, so that $\meff^2 = 2H^2$.
The conformal factor in the metric is absorbed via a field redefinition
\cite{ratra}.
We follow Ref.\ \cite{allen} in our analysis and work to leading order in
the expansion of $\tphi(x)$.
Anticipating our discussion of a dynamical $\al$, we shall treat $\al(x-y)$ as an
object with the mass dimension of a canonical scalar.
This implies that the parameter $\lambda$ in the definition of $\Sal$ has 
dimension $({\rm mass})^2$.\footnote{In general, $\lambda$ has dimension $({\rm mass})^{d/2}$.} 
As the Hubble parameter is the only natural scale in the theory, we take $\lam \simeq H^2$.
The modes are
\be
\phi_{\kk}(\tau,\x) = \frac12 H \tau^{3/2} H_\nu^{(2)}(k\tau) e^{i\kk\cdot\x}, \label{mexp}
\ee
where $k = (\kk\cdot\kk)^{1/2}$ and $\nu = (\frac94 - \meff^2 H^{-2})^{1/2} = \frac12$. 
Working with the mode expansion of $\tphi(x)$ to leading order in the expansion
parameter $\lam\, \al(k)$, one finds 
\bea
\tphi(x) &=& \phi(x) + \lam \int d^3y\ \al(\y_A - \x) \phi(y) + \ldots \label{ps} \\
&\simeq& \int \frac{d^3k}{(2\pi)^3} H\tau^{3/2} \left\{ 1 + \lam \al(k) + \ldots \right\}
\left[ H_\nu^{(2)}(k\tau) e^{i\kk\cdot\x} a_{\kk} + H_\nu^{(1)}(k\tau) e^{-i\kk\cdot\x} a_{\kk}^\dagger \right].
\eea
The integral \eref{ps} is strictly defined on the $t=0$ surface in global coordinates. 
In practice, we substitute in the mode expansion \eref{mexp} and Fourier transform.
The expression in curly brackets determines the corrections to the power spectrum in this admittedly na\"{\i}ve scheme:
\be
\widetilde{P}(k) = \left( 1 + 2\lam\al(k) + \ldots \right) P(k).
\ee
We find that
there is a momentum dependent scaling in the power spectrum that arises from looking at the autocorrelation
function of $\tphi$ in the Euclidean vacuum.
In four dimensions, a normalizable $\al$-state obtains from $\al(k) \sim k^{-2}$ scaling.
For such a profile, we expect $\CO((H/k)^2)$ deviations from correlated initial conditions. 

The correction to the power spectrum that we compute depends explicitly on the momentum.
It is determined by the profile of the squeezed state $\al(k)$.
The $\al$-vacuum and adiabatic vacuum calculations of Refs.\ \cite{ulf, stanford} predict either $\CO(H/\om)$ or
$\CO((H/\om)^2)$ deviations as signals of trans-Planckian physics.
The scale $\om$ is an ultraviolet cutoff,
which is fixed.
There is no explicit momentum dependence in the correction to the power spectrum.
Ref.\ \cite{ulf} computes the power spectrum to be
\be
\widetilde{P}(k) = \left( \frac{H}{2\pi} \right)^2 \left( 1 - \frac{H}{\om} \sin\left( \frac{2\om}{H} \right) \right).
\ee
This should correspond to $\al(k) \simeq 1/k$, which is not normalizable as an excitation over the Euclidean vacuum. 
The claim is that this details the physics of an $\alpha$-vacuum.
Using effective field theory in the Euclidean vacuum, Ref.\ \cite{stanford} corrects the power spectrum as
\be
\widetilde{P}(k) = \left( \frac{H}{2\pi} \right)^2 \left( 1 + \chi \left(\frac{H}{\om}\right)^2 \right),
\ee
where $\chi$ is a model dependent numerical factor.

Kaloper and Kaplinghat \cite{kk} consider a sudden change in the background during the late stages of inflation.
The inflaton, which defines the ground state of the quantum system prior to the sudden transition, 
is trapped in a squeezed state above the adiabatic vacuum after the transition.
Because the inflaton is in an excited state, the power spectrum will deviate from the standard thermal result.
Ref.\ \cite{kk} computes the corrections to go as
\bea
\widetilde{P}(k) &=& \left( 1 + \CD(p, H, \epsilon_H, \eta_H) \right) P(k), \\ 
\CD(p, H, \epsilon_H, \eta_H) &=& 
\Phi(\epsilon_H, \eta_H) +
\left( \frac{H}{p} \right)^2 \cos\left( \frac{2p}{H} \right) +
\Delta(\epsilon_H, \eta_H) \frac{H}{p} \sin\left( \frac{2p}{H} \right),
\label{kkps}
\eea
where $\Phi$ and $\Delta$ are functions of the slow-roll parameters and $p=k/a_0$.
If we expand out the trigonometric functions in eq.\ \eref{kkps}, we see that the leading order correction in the
power spectrum $\Delta P(k) \simeq (H/k)^2 P(k)$, in accord with our computations. 

Initial conditions corresponding to squeezed excitations are imprinted in the CMBR.
It is conceivable that deviations from the standard inflationary predictions can be detected in future observations.
In principle, the $\al$-state will affect higher order correlation functions as well
as the two-point functions.
(Three-point correlation functions for scalar fields in inflationary models were studied
in Ref.\ \cite{malda}.) 

\subsection{$\al$ as a dynamical field}

We know that in QCD the $\theta$-angle, which is a superselection parameter,
can be promoted to a dynamical field, the axion, consistent with
phenomenological constraints on CP violation in the strong sector \cite{theta}.
The axion is the pseudoscalar Goldstone boson associated to a spontaneously
broken global chiral $U(1)$ symmetry (the Peccei-Quinn symmetry).
In an interacting scalar field theory in de Sitter space, quantum effects
in the infrared {\em restore} spontaneously broken symmetries \cite{ratra}.
Giving $\alpha$ dynamics in de Sitter space must proceed in a different way.

We can approach the problem from a number of perspectives:
Firstly,
the dynamics of $\alpha$ may be induced by integrating out other fields to
which $\phi(x)$ couples.
As the $\phi$--$\alpha$ system evolves in an inflationary cosmology, dynamics
can then dump the energy from one field into the other so that at late times we
end up in the Euclidean vacuum.
Secondly,
the dynamics of $\alpha$ may be induced by self-interactions of $\phi(x)$.
The self-interactions can then be replaced by coupling to an auxiliary field $\alpha$,
just as one integrates in a new degree of freedom.
Thirdly,
as the antipodal map invokes $\al(x-y)$, the translational invariance may
suggest that $\al$ is a condensate rather than a fundamental field:
$\al(x-y) = \vev{\overline\psi(x)\psi(y)}$.
The initial conditions in each case must be fine tuned to ensure the proper entanglement
between $\phi(x)$ and $\phi(x_A)$.
As {\em any} particular initial condition {\em including the Euclidean one} is finely tuned,
this is not a criticism unique to this model.
It is, however, a challenge to obtain the particular excited states that we are interested in exploring
as an effective field theory. 

As a stepping stone to promoting the $\alpha$ superselection parameter to a fully
dynamical field, the {\em alphon}, let us instead say a few words about the more
modest effort of endowing the $\al$-state with dynamics.
To the Hamiltonian density $\widehat\CH$ from eq.\ \eref{ch}, we append
\be
\widehat\CH_\al = \frac12 \pi_\al^2 + \frac12 (\nabla\al)^2 + V(\al).
\ee
We treat $\al$ as an additional scalar field in the theory, with a canonical $[\al,\pi_\al]$
commutator and canonical mass dimension.
This field couples to $\phi$ through the definition of $\tphi$.
The field $\tphi$ is an excitation over the dynamically evolving state $\ket\al$.

We assume that $\al$ and $\pi_\al$ commute with $\phi$ and $\pi$.
This immediately implies that $\al$ commutes with $\tphi$ and $\tpi$.
However, $\pi_\al$ {\em will not commute} with either $\tphi$ or $\tpi$ because
factors of $\al$ appear explicitly in the field redefinitions
\eref{tphi} and \eref{tpi}.
We have
\bea
&& [\tphi(x), \pi_\al(x')] = 
i\lam\, \tphi(x_A+x'_A) + \CO(\lam^2),
\\ 
&& [\tpi(x), \pi_\al(x')] = 
-i\lam\, \tpi(x_A-x') + \CO(\lam^2).
\eea
This will lead to a coupled set of non-local integro-differential equations
for the combined $\tphi$--$\al$ system.

Although it is difficult to solve these equations even perturbatively in the parameter $\lam$ in low dimensions,
we nevertheless expect that a solution exists and is stable.
Our intuition for this arises from Minkowski space, where we can construct $\al$-like squeezed states and
consider excitations over them.
The only difference in the formul\ae\ \eref{Sal}, \eref{tphi}, \eref{tpi} is that we remove the
antipodal labels.
Promoting $\al$ to a dynamical variable in Minkowski space results in a similar set of coupled
non-local integro-differential equations.
As we expect the physics of such highly correlated entangled states to make perfect sense in 
flat Minkowski space, we expect nothing different in de Sitter space.
The only new wrinkle in de Sitter space is the presence of antipodal labels on certain coordinates, but the
structure of the equations of motion are the same.
Indeed, from the point of view of the inflationary patch in de Sitter space where there is no
clear notion of what the antipode is, Minkowski space reasoning is perhaps the most appropriate
framework in which to address physics.

\paragraph{The rolling vacuum --- a toy model} ${}$ \\
To investigate to what extent a slow-rolling alphon keeps the field
$\tphi(x)$ in the $\alpha$-vacuum, we consider a similar problem for
squeezed states in a harmonic oscillator. Thus, as in
Section \ref{sec:opal}, we take the operator
\be
\CU_\al=\exp\left[\frac{1}{2}\, \al \left((a^{\dagger})^2 - a^2\right)\right],
\ee
One easily verifies that $\CU_\al$ creates
a squeezed state when acting on the ground state of the harmonic
oscillator,
\be
\CU_\al\ket{0}= \frac{1}{\sqrt{\cosh\alpha}} \exp\left[\frac{1}{2} 
\tanh\alpha \, (a^{\dagger})^2 \right] \ket{0}.
\ee
We now construct a Hamiltonian $\Ham_\al$, whose ground state is 
$\CU_\al\ket{0}$. This Hamiltonian is the analogue of the alphon
Hamiltonian constructed before, and it is simply given by
\be \Ham_\al=\CU_\al \Ham \CU_\al^{\dagger},
\ee
with $\Ham$ the usual harmonic oscillator Hamiltonian. The exact eigenstates
of $\Ham_\al$ are clearly the states $\CU_\al\ket{n}$, with $\ket{n}$ the
usual harmonic oscillator states $\ket{n}=(n!)^{-1/2} (a^{\dagger})^n\ket{0}$.
Next, we are going to make $\al$ time-dependent, and study the corresponding
Schr\"odinger equation
\be
{i \hbar} \frac{\partial\psi}{\partial t} = \CU_{\al(t)} \Ham \CU_{\al(t)}^{\dagger} \psi(t).
\ee
Assuming that there is no backreaction on $\alpha$, a general solution will be of the form
\be
\psi(t) = \sum_{k=0}^{\infty} c_k(t) e^{-i E_k t/\hbar} \CU_{\al(t)}\ket{k}.
\ee
To get an idea of the time-dependence, we take $\al(t)=\al_0 \exp(-\epsilon t)$,
and start with the state $\CU_{\al_0}\ket{0}$ at $t=0$. To leading
order in $\epsilon$, we find that the leakage out of the ground state is
\be \label{defdelta}
\Delta \equiv \sum_{\ell>0} |c_\ell(t)|^2 = 
\frac{1}{8} \left| \frac{\al_0 \epsilon}{\omega} \right|^2.
\ee

For a massive field in flat space, one can directly apply a similar
calculation, with $\omega$ replaced by $\sqrt{k^2 + m^2}$, with $k$
the spatial momentum and $m$ the mass of the field. The parameter
$\al(t)$ can now also be a non-trivial function of the momentum $k$,
$\al(t) \rightarrow \al(k,t)$.
In $d$-dimensional de Sitter space this situation is somewhat different. 
For a massive field with sufficiently large mass, $m\geq (d-1)/2$,
the early and late
time behavior of the solutions to the equations of motion is of the form $t^{\pm (d-1)/2
\pm i\mu}$ with $\mu^2=m^2-(d-1)^2/4$. In particular, this does not
depend on the momentum $k$ of the mode under consideration. In contrast
to what happened in flat space, we should replace $|\omega|$ by $m$,
not by $\sqrt{k^2+m^2}$. At early times, in the inflationary patch,
the field shows the usual oscillatory behavior, and the transition
between the two behaviors takes place at time $t^{\ast}\sim \log k$.

To summarize, we reinstate the scale $M$, which is of order the
Hubble parameter $H$. 
The metric for de Sitter space is
\be
ds^2 = -dt^2 + e^{2Ht} dx^2.
\ee
Then as long as we consider a mode with momentum mode $k$ at times
$t$ which satisfy $tM \gg \log(k/M)$, the deviation from the vacuum
$\Delta$ defined in eq.\ (\ref{defdelta}) generated by alphon decay 
is of the order of
\be 
\Delta \sim \left( \frac{\dot{\alpha}(k)}{m/M} \right)^2.
\label{ack}
\ee
%

\paragraph{Analogy with decoherence} ${}$ \\
Perhaps one route to understanding the behavior of $\al$ is to proceed in
analogy with decoherence and regard the 
combined $\phi$--$\alpha$ system as a harmonic oscillator $\alpha$ coupled to
an environment $\phi$ via a frictional term.
The r\^{o}le of dissipation in such quantum systems has been addressed by
Feynman and Vernon \cite{fv} and Caldeira and Leggett \cite{cl}.
Let us consider the following Hamiltonian:
\be
\Ham = \frac12 m\dot{x}^2 + \frac12 m \om^2 x^2 + \frac12 M\dot{q}^2 + V(q) - \gamma x q.
\ee
The oscillator $q$ is linearly coupled to the environment $x$ by a damping term.
To deduce the dynamics of $q$, one integrates out $x$ as follows:
\bea
F_1[f;x,x'] &=& 
\int [{\cal D}x]\ \exp\left[-\frac1\hbar \int_0^T dt\ \left(\frac12 m\dot{x}^2 + \frac12 m\om^2 x^2 + i f(t) x(t)\right)
\right]_{x(0)=x, \ x(T)=x'} \\
&=& \left(\frac{m\om}{2\pi\hbar} \sinh(\om T)\right)^{1/2} e^{-\Phi/\hbar}, \\
\Phi &=& \frac{1}{4m\om} \int_0^T dt\, \int_0^T dt'\ e^{-\om|t-t'|} f(t) f(t') + \frac{m\om}{2\sinh(\om T)} \times \\
\nn && [(x^2 + x'^2) \cosh(\om T) - 2 x x' + 2 A(x e^{\om T} - x') + 2 B(x' e^{\om T} - x) + (A^2 + B^2) e^{\om T} - 2 A B], \\
A &=& \frac{i}{2m\om} \int_0^T dt\ e^{-\om t} f(t), \\
B &=& \frac{i}{2m\om} \int_0^T dt\ e^{-\om(T-t)} f(t).
\eea
Then
\bea
Z[f] &=& \left\langle exp\left[-\frac{i}\hbar \int_0^T dt\ f(t) x(t) \right] \right\rangle_{\rm harm.\ osc.} \\
&=& \frac{\int dx\ F_1(f;x,x)}{\int dx\ F_1(0;x,x)} \\
&=& \exp\left[-\frac{1}{4m\om\hbar} \int_0^T dt\, \int_0^T dt'\ \frac{\cosh(\om|t-t'|-\frac12\om T)}{\sinh(\frac12\om T)}f(t) f(t')\right].
\eea
Taking $q(t)$ to be periodic with period $T$, we obtain the effective action 
\bea
S_{\rm eff}[q] &=& \int_0^T dt\ \left[\frac12 M\dot{q}^2 + V(q)\right] - 
                   \int_{-\infty}^{\infty} dt\, \int_0^T dt'\ A(t-t') q(t) q(t') + {\rm const.}, \\
A(t-t') &=& \frac{\gamma^2}{4m\om} e^{-\om|t-t'|} =: \frac{1}{2\pi} \int d\om\ J(\om) e^{-\om|t-t'|}.
\eea

In the Feynman-Vernon formalism, the coupling to the environment is always linear,
whereas
at leading order, the interaction between $\alpha$ and $\phi$ will include terms like 
\be
\Ham_{\rm int} \supset \frac12 \lam\, \meff^2\, \int d^{d-1}x\, d^{d-1}y\, \alpha(y-x) \phi(x) \phi(y_A) + \ldots.
\ee
This differs in many respects from the dissipation analysis above, which is purely Gaussian.
Nevertheless, pursuing this reasoning further may offer insight into the dynamics of $\alpha$. 

\section{Infrared Corrections to Gravity?}
\label{sec:algr}

As we have noted earlier, it is possible to extend our definition
of the antipodal map to fields of other spin.
The extension to vector fields is straightforward: we replace $\phi(x_A)$
with $A_\mu(x_A)$ and $\pi(y)$ with $\pi_A^\mu(y) = \frac{-i}{\sqrt{-g}}\frac{\delta}{\delta A_\mu(y)}$
in eq.\ \eref{Sal}.
The extension to fermions is complicated only in that we must preserve
an anticommutator instead of a commutator in the canonical transformation.
The Hamiltonian transforms by conjugation as
\bea
e^{i\Sal} \cdot i \psi^\dagger(x) \pa_t\psi(x) \cdot e^{-i\Sal}
&=& i e^{i\Sal} \cdot \psi^\dagger(x) \cdot e^{-i\Sal} \cdot e^{i\Sal} \pa_t\psi(x) \cdot e^{-i\Sal} \nn \\
&=:& i \widetilde\psi^\dagger(x) \pa_t\widetilde\psi(x).
\label{spinors}
\eea 
We can check that under the field redefinition of eq.\ \eref{spinors},
$\{\widetilde\psi(x), \widetilde\pi_\psi(x')\} = \{\psi(x), \pi_\psi(x')\}$.
(Recall that $\pi_\psi(x) = i\psi^\dagger(x)$.)

Considering an antipodal map for spin-2 fields, the metric in particular,
is more interesting.

It has been suggested in a number of contexts \cite{nima} that
infrared modifications to gravity may play an important r\^{o}le in
elucidating the physics of the vacuum, namely in addressing the cosmological
constant problems.
Can we utilize the formalism we have developed to address matters of gravity?

We define the conjugate momentum to the metric as
$\pig{\mu\nu}(t,x) = \frac{-i}{\sqrtg{t,x}}\, \frac{\delta}{\delta \g{\mu\nu}(t,x)}$.
Note that the canonical equal time commutator is
\be
[\g{\mu\nu}(t,x), \pig{\rho\sigma}(t,x')] 
= i\delta_\mu^\rho \delta_\nu^\sigma \frac{\delta^{(d-1)}(x-x')}{\sqrtg{t,x'}} \label{comm}.
\ee
We take
\be
\Sal = \lambda \int d^{d-1}x\, \sqrtga{x}\, \int d^{d-1}y\, \sqrtga{y}\, 
       \al(x-y) \g{\mu\nu}(t=0,x_A) \pig{\mu\nu}(t=0,y) \label{Salg}.
\ee
The integrals are defined on the $\Sigma_{t=0}$ surface as before.
As the de Sitter line element in global coordinates is
\be
ds^2 = \g{\mu\nu}(t,x) dx^\mu dx^\nu = -dt^2 + H^{-2}(\cosh^2 Ht)d\Omega_{d-1}^2,
\ee
$\left. \sqrtg{t,x} \right|_{t=0} = \sqrtga{x}$.
(We have chosen a gauge with lapse set to unity and shift set to zero.)
However, the formul\ae\ we obtain for
\bea
\tg{\mu\nu}(x) &=& e^{i\Sal} \g{\mu\nu}(x) e^{-i\Sal}, \\
\tpig{\mu\nu}(x) &=& e^{i\Sal} \pig{\mu\nu}(x) e^{-i\Sal},
\eea
are complicated by the action of $\delta/\delta g_{\mu\nu}$ on the
various determinant factors.

Conjugation by the antipodal map gives
\bea
\tg{\mu\nu}(0,x) &=& \g{\mu\nu}(0,x) + [i\Sal, \g{\mu\nu}(0,x)] + \frac{1}{2} [i\Sal, [i\Sal, \g{\mu\nu}(0,x)]] + \ldots \\
&=& \g{\mu\nu}(0,x) + \lambda \int d^{d-1}y\, \sqrtga{y}\, \al(y-x) \g{\mu\nu}(0,y_A) \nonumber \\
&& + \frac{1}{2} \lambda^2 \int d^{d-1}y\, \sqrtga{y}\, \int d^{d-1}z\, \sqrtga{z}\, 
     \al(y-x) \al(z-y_A) \g{\mu\nu}(0,z_A) \\
&& + \frac{1}{4} \lambda^2 \int d^{d-1}y\, \sqrtga{y}\, \int d^{d-1}z\, \sqrtga{z}\, 
     \al(y-x) \al(z-y) \g{\mu\nu}(0,y_A) g^{\rho\sigma}(0,y) \g{\rho\sigma}(0,z_A) \nonumber \\
&& + \CO(\lambda^3). \nonumber
\eea
Starting at $\CO(\lambda^2)$, we find terms in $\tg{\mu\nu}$ from the action
of derivatives on the measure. 

Let us work in the weak field approximation by treating $\lambda\ \alpha(k)$ as a
small parameter.  
We take
\bea
\tg{\mu\nu}(0,x) = \g{\mu\nu}(0,x) + h_{\mu\nu}(0,x), &&
h_{\mu\nu}(0,x) = [i\Sal, \g{\mu\nu}(0,x)].
\eea
Indices are lowered and raised using the background (inverse) metric
$\g{\mu\nu}$ and $g^{\mu\nu}$.
We define $h := g^{\mu\nu} h_{\mu\nu}$.
In this linearized gravity approximation
\bea
&& \widetilde{g}^{\mu\nu} = g^{\mu\nu} - h^{\mu\nu} + h^{\mu\rho} h_\rho^\nu + \ldots, \\ 
&& \sqrt{-\widetilde{g}} = \sqrt{-g} \left(1 + \frac12 h + \frac18 h^2 - \frac14 h^{\mu\nu} h_{\mu\nu} + \ldots \right).
\eea
As $h_{\mu\nu}$ is itself proportional to $\lambda$, we shall drop
quadratic and higher order terms in $h$.
 
At $\CO(\lambda)$, we find that
\bea
\tpig{\mu\nu}(0,x) &=& \pig{\mu\nu}(0,x) - 
\lambda \frac{\sqrtga{x_A}}{\sqrtga{x}} 
\int d^{d-1}y\, \sqrtga{y}\, \al(x_A-y) \pig{\mu\nu}(0,y) \nonumber \\
&& - \frac{1}{2} \lambda g^{\rho\sigma}(0,x)
\int d^{d-1}y\, \sqrtga{y}\, \al(y-x) \g{\rho\sigma}(0,y_A) \pig{\mu\nu}(0,x) \\
&& - \frac{1}{2} \lambda g^{\mu\nu}(0,x) \g{\rho\sigma}(0,x_A) 
\int d^{d-1}y\, \sqrtga{y}\, \al(x-y) \pig{\rho\sigma}(0,y) 
+ \CO(\lambda^2). \nonumber
\eea
A careful computation then gives 
\bea
[\tg{\mu\nu}(0,x), \tpig{\rho\sigma}(0,x')] &=&  
[\g{\mu\nu}(0,x), \pig{\rho\sigma}(0,x')] -
\frac{i}{2} \delta_\mu^\rho \delta_\nu^\sigma \frac{\delta^{(d-1)}(x-x')}{\sqrtg{0,x'}}\, 
g^{\lambda\tau}(0,x') [i\Sal, \g{\lambda\tau}(0,x')] + \ldots \nn \\ && \\ 
&=& i \delta_\mu^\rho \delta_\nu^\sigma \frac{\delta^{(d-1)}(x-x')}{\sqrtg{0,x'}} \left(1 - \frac12\, h(0,x') + \ldots \right) \\
&=& i \delta_\mu^\rho \delta_\nu^\sigma \frac{\delta^{(d-1)}(x-x')}{\sqrt{-\widetilde{g}(0,x')}}. 
\eea
This is of course the expected result if we act by conjugation on the right hand side of
eq.\ \eref{comm}, but it is not a canonical transformation in that commutation
relation for the metric and its conjugate momentum is not preserved.\footnote{
To make a canonical transformation so that
\be
[\tg{\mu\nu}(0,x), \tpig{\rho\sigma}(0,x')] = [\g{\mu\nu}(0,x), \pig{\rho\sigma}(0,x')] \nn
= i \delta_\mu^\rho \delta_\nu^\sigma \delta^{(d-1)}(x-x'),
\ee
one can instead work with the momentum density, 
$\pi_g^{\mu\nu}(t,x) = -i \frac{\delta}{\delta g_{\mu\nu}(t,x)}$.
We have
\bea
\tg{\mu\nu}(0,x) &=& g_{\mu\nu}(0,x) + h_{\mu\nu}(0,x) + \ldots, \;\;\;\;\; 
h_{\mu\nu}(0,x) = [i\Sal, g_{\mu\nu}(0,x)] = \lam \int d^{d-1}y\, \sqrt{\gamma(y)}\, \al(y-x) g_{\mu\nu}(0,y_A), \nn \\
\tpig{\mu\nu}(0,x) &=& \pi_g^{\mu\nu}(0,x) - \lam \sqrt{\gamma(x_A)}\, \int d^{d-1}y\, \sqrt{\gamma(y)}\, \al(x_A-y) \pi_g^{\mu\nu}(0,y) \nn \\
&& -\frac12 \lam \sqrt{\gamma(x)}\, g^{\mu\nu}(0,x) \int d^{d-1}y\, \sqrt{\gamma(y)}\, 
   \left[ \al(y-x) g_{\rho\sigma}(0,y_A) \pi_g^{\rho\sigma}(0,x) + \al(x-y) g_{\rho\sigma}(0,x_A) \pi_g^{\rho\sigma}(0,y) \right] + \ldots. \nn 
\eea
}

%

Working with the new metric $\tg{\mu\nu}$, we can compute the change in the curvature:
\bea
\widetilde{R}_{\mu\nu}[g,h] &=& R_{\mu\nu}[g] + \delta R_{\mu\nu}[g,h], \\
\delta R_{\mu\nu}[g,h] &=& 
\frac12 \left[ \nabla_\lambda \nabla_\mu h_\nu^\lambda +
               \nabla_\lambda \nabla_\nu h_\mu^\lambda -
               \nabla_\lambda \nabla^\lambda h_{\mu\nu} -
               \nabla_\mu \nabla_\nu h \right],
\eea 
where $\nabla_\mu$ is the connection with respect to the fixed background metric $\g{\mu\nu}$.
Since $\nabla_\mu$ is a covariant derivative with respect to $x$, it acts on the factors of 
$\alpha$ in the integral expression for $h_{\mu\nu}$.
The Einstein equation gives
\be
\delta G_{\mu\nu} := \delta R_{\mu\nu} - \frac12 R\, h_{\mu\nu} -\frac12 \delta R\, \g{\mu\nu} + \Lambda\, h_{\mu\nu} = 8\pi G_N T_{\mu\nu}.
\ee
Writing this explicitly in terms of $\al$ is not especially illuminating.
We observe though that the effect of conjugation by the antipodal map can be absorbed into a modification of the stress-energy.
Once we choose an $\al$-state, the gravitational interaction --- \ie the way geometry couples to matter --- involves the Hubble scale.
This is an infrared effect in gravitational physics that can be treated perturbatively in the $\al$-expansion of the antipodal map.
Perhaps a detailed analysis of such modifications will reveal interesting phenomenology.
A second phenomenological manifestation of $\al$ in the gravity sector may appear in deviations from the standard power spectrum for tensor fluctuations.

\section{The Elegant de Sitter Universe}
\label{sec:conc}

The low energy effective dynamics of gravitation as a classical field theory and the physics of the Standard
Model as a quantum field theory that captures a specific set of gauge principles is expected to descend from
a consistent theory of quantum gravity. 
String theory is a very promising candidate theory of quantum gravity and matter.
Among the many outstanding problems of string theory is the absence of a vacuum selection principle 
that marries with empirical data about the Universe.
Our intuition for string theory/quantum gravity, and indeed our calculational technology, is essentially rooted in
field theory.
Perturbative string theory tells us how to compute $S$-matrix observables (scattering amplitudes) in ten-dimensional
flat spacetime.
The notion of well defined particle excitations at asymptopia is, however, an artifact of the flat space limit.
Though the classic techniques of field theory are the tools we rely upon to make sense of the initial singularity and 
the emergence of space, time, and matter, it is not clear that the tools we have are appropriate to answer the
questions we ask.
In de Sitter space there is no $S$-matrix.
Different local observers have different horizons and make different measurements.

As well, the requirements of spacetime supersymmetry are incompatible with the symmetries of de Sitter space.
There is no positive conserved energy because there is no global time-like Killing isometry. 
To date, there has been no construction of de Sitter space as a consistent background geometry for superstrings.
To realize de Sitter space in string theory, one may consider, for instance, Type IIB compactifications with a suitable choice
of background flux \cite{kklt}.
Such de Sitter vacua are at best metastable and finely tuned.
The true vacuum is ten-dimensional flat space.

Quantum field theory in de Sitter space, on the other hand,
does make sense at least as an algebraic theory \cite{wald} and provides a 
framework to address a number of the issues that persist in quantum gravity.
As de Sitter space admits an infinite family of vacua, we need a vacuum selection principle.
One particular vacuum of this family is uniquely consistent with the Minkowski space limit of the de Sitter geometry.
We have seen that squeezed states in the Fock space constructed from this unique vacuum share many of the features of
the other de Sitter $\al$-vacua.
The functional formalism provides a crisp technology for addressing the implications of initial conditions with
the long-range correlations native to these $\al$-states.
The effects of backreaction on $\alpha$ may modify some of this physics.

Initial conditions for inflation are generic in the sense that many sets of inputs lead to the same large scale
structures today in causally disconnected regions of spacetime.
Nevertheless some details about the initial conditions can in principle be imprinted in the spectrum of the CMBR.
If the Universe were to have begun in an $\al$-state, the power spectrum for the inflaton may exhibit a novel
dependence on momentum. 

More radically, we can imagine that the $\al$ parameter that determines the vacuum and the Hilbert space is a fully
dynamical superselection parameter, just as the $\theta$ parameter of QCD becomes the fully dynamical axion.
In such a setting, what we mean by ``vacuum'' dynamically rolls toward the Euclidean value as the Universe 
inflates and drags the vacua of the other fields along with it.
There are obvious phenomenological questions to address in this scenario.
For example, is it consistent with the standard pictures of inflation, reheating, and Big Bang nucleosynthesis? 

Recently it has been suggested that dynamics selects a corner of the moduli space of supersymmetric vacua of string
theory (the so-called {\em landscape of vacua}) consistent with measurements of (some set of) physical parameters in
our Universe \cite{anthro}.
The cosmological constant, for example, is its measured value in the corner of moduli space where our Universe resides.
Superselection parameters, like $\theta$ of QCD or $\alpha$ for fields in de Sitter space (it is worth repeating here
that each of the Standard Model degrees of freedom will have an associated $\alpha$) will vary throughout the 
landscape just as the cosmological constant $\Lambda$ does.
The requirements of 
large scale structure in the Universe,
inflation, and 
compatibility with Big Bang nucleosynthesis
do not place restrictions on the numerical values of $\theta$ or $\alpha$.
Rather, like the existence of a third generation of Standard Model matter fields, the details of the inflationary
power spectrum that is shaped by the $\alpha$-state is an additional free parameter in the description of the landscape. 
It seems that this condition is generic for superselection parameters in quantum gravity.
Perhaps this places additional limits on the predictivity of a theory of physics that includes anthropics as a defining
principle.

It has been argued in Ref.\ \cite{bms} that the one-parameter family of de Sitter invariant $\alpha$-vacua can be 
understood as a one-parameter family of marginal deformations of a possible CFT dual to the bulk de Sitter space.
This interpretation is quite natural in view of the fact that the $\alpha$ parameter appears as a superselection
parameter in the general de Sitter invariant wavefunctional.
Alternatively, it was noticed that entangled states preserving $SO(1,d)$ invariance in the product of two CFTs of the
type discussed in Ref.\ \cite{bdm2} closely resemble boundary states for a free scalar field, for which a one-parameter
ambiguity has been known in the literature.
It was thus suggested that this translates into a one-parameter ambiguity of the de Sitter invariant vacua.
It would be interesting to understand $\al$-states in de Sitter space from the point of view of the conjectured 
de Sitter/CFT correspondence.
The thermofield picture of evolution off the $\Sigma_{t=0}$ hypersurface on which the antipodal transformation is made
may improve our knowledge about how boundary data on $\CI^\pm$ become entangled.

Finally, we have seen that the formalism for scalar fields extends to metric fluctuations.
Hubble scale correlations in the geometry lead to infrared modifications of the gravitational interaction.
This is a rich and promising direction with phenomenology to explore in greater depth.

We hope that continued analysis of physics in de Sitter space will shed new light on outstanding issues in cosmology
and quantum field theory.
Perhaps we are fortunate to live in an elegant de Sitter Universe. 

\vskip 1.5 cm
\section*{Acknowledgments}

We owe much of our understanding of the physics of de Sitter space to Vijay Balasubramanian.
It is a pleasure to thank him for his collaboration and insights at various stages of this 
work.
We have benefited as well from conversations with many colleagues including
Tom Banks,
Willy Fischler,
Esko Keski-Vakkuri, 
Nemanja Kaloper, 
Matt Kleban, 
Per Kraus, 
Finn Larsen, 
Rob Leigh, 
Thomas Levi, 
Alex Maloney, 
Asad Naqvi, 
Simon Ross, 
Koenraad Schalm,
Steve Shenker, and
Jan-Pieter van der Schaar.
JdB and DM were supported in part by NSF grant PHY-9907949 at the KITP, Santa Barbara.
VJ and DM were supported in part by the U.S.\ Department of Energy under contract DE-FG05-92ER40709.
VJ thanks the High Energy Group at the University of Pennsylvania for their generous hospitality.
Finally, JdB, VJ, and DM acknowledge the hospitality and support of the Aspen Center for Physics.

\end{document}